\documentclass[3p,times]{elsarticle}

\usepackage{ecrc}

\usepackage{epstopdf}
\usepackage{subfig}

\volume{00}

\firstpage{1}

\journalname{Nuclear Physics A}

\runauth{C. Andrei et al.}

\jid{nupha}

\jnltitlelogo{Nuclear Physics A}

\usepackage{graphicx}
\usepackage{amsmath,amssymb}

\newcommand{\pbar}{$\rm\overline{p}$}
\newcommand{\pt}{\ensuremath{p_{\rm T}}}
\newcommand{\s}{$\sqrt{s}$}
\newcommand{\pip}{$\pi^{+}$}
\newcommand{\kap}{K$^{+}$}

\begin{document}

\begin{frontmatter}



\title{Light flavor hadron spectra at low \pt{} and search for collective phenomena in high multiplicity pp, p--Pb and Pb--Pb collisions measured with the ALICE experiment}

\author{C. Andrei (for the ALICE Collaboration)}
\address{National Institute for Physics and Nuclear Engineering, Bucharest, Romania}




\begin{abstract}
Comprehensive results on transverse momentum distributions and their ratios for identified light flavor hadrons ($\pi$, K, p) at low $p_{T}$  and mid-rapidity as a function of charged particle multiplicity are reported for pp collisions at 7 TeV. 
Particle mass dependent hardening of the spectral shapes in Pb--Pb collisions at 2.76 TeV were attributed to hydrodynamical flow and quantitatively parameterized with Boltzmann-Gibbs Blast Wave fits. 
In this contribution, we investigate the existence of collective phenomena in small systems: pp, p--Pb and peripheral Pb--Pb where similar patterns are observed in multiplicity dependent studies.
\end{abstract}

\begin{keyword}
LHC \sep ALICE \sep light-flavor hadrons \sep multiplicity \sep transverse momentum spectra \sep collective phenomena 

\end{keyword}

\end{frontmatter}



\vspace{1cm}
Identified charged particle transverse momentum spectra in p--pbar  collisions were extensively studied as a function of incident energy below 900 GeV at CERN Sp\pbar S \cite{ansorge} and up to 1800 GeV at Fermilab Tevatron \cite{alexopo1}. 
Starting from 200 GeV, a deviation of the $\langle p_{T} \rangle$ of kaons as a function of \s{} relative to the expectations based on the extrapolation of ISR data was reported by the UA5 Collaboration \cite{alner}. 
The E735 Collaboration saw evidence for a mass dependence of the $\langle p_{T} \rangle$ as a function of c.m. energy from 300 to 1800 GeV and of the $\langle p_{T} \rangle$ as a function of d$N_{ch}/{\rm d}\eta$ at 1800~GeV \cite{alexopo1}.
The origin of these experimental observations is still under debate.
QCD inspired models like PYTHIA~\cite{sjos} and EPOS \cite{werner} reproduce the experimental trends observed at Tevatron considering multiple partonic interactions and rescattering or a hydrodynamic type evolution with flux tube initial conditions, respectively.  
At about four times larger incident energies, the case of the present study, such processes become more important and contribute to a large energy transfer, well beyond the deconfinement energy density.  
Therefore, it is quite probable that at such energies a piece of matter of proton size explodes hydrodynamically once the energy transfer becomes significantly large \cite{bel}.

Transverse momentum distributions and their relative ratios for identified positive charged hadrons \pip{}, \kap{}, p at mid-rapidity in pp collisions at $\sqrt{s}$ = 7 TeV, measured by the ALICE experiment \cite{ali1} at the LHC, as a function of charged particle multiplicity are obtained.
The \pt{} range goes from 0.2 GeV/\textit{c}, 0.3 GeV/\textit{c} and  0.5 GeV/\textit{c} to 2.6 GeV/\textit{c}, 1.4 GeV/\textit{c} and 2.6 GeV/\textit{c} for \pip{}, \kap{} and p respectively, significantly larger than previously reported results \cite{CMS}.
60 million inelastic pp collisions collected by the ALICE experiment during the 2010 run at LHC, using minimum bias (MB) and high multiplicity (HM) triggers \cite{performance}, at \s{} = 7 TeV, were analyzed.
Events corresponding to the measured charged particle multiplicity within the pseudorapidity range $|\eta|<$ 0.8 were selected.
The multiplicity was estimated summing the global tracks and complementary Inner Tracking System (ITS) standalone tracks and Silicon Pixel Detector (SPD) tracklets. 
Events were required to have a primary vertex within $\pm$10 cm from the center of the Time Projection Chamber (TPC) and the Time-Of-Flight array (TOF) subdetectors.
A \pt{} dependent Distance of Closest Approach (DCA) cut of (0.018 + 0.035$p_T^{-1.01}$) cm in the transverse plane and 2 cm in the longitudinal direction were used in order to reduce the contamination from weak decays of strange particles, conversion or secondary hadronic interactions in the detector material.
The present analysis has been done in $|y|$ \textless{} 0.5.
The particle identification was based on a Bayesian approach using the information from ITS, TPC and TOF. 
The detector response functions and the \textit{a priori} probabilities were obtained from the experimental data.
The contaminations from weak decays products and from particles produced by the interactions with the detector material were estimated based on a data driven method. 
\begin{figure}
\begin{center}
\subfloat[]{\includegraphics*[width=0.51\linewidth]{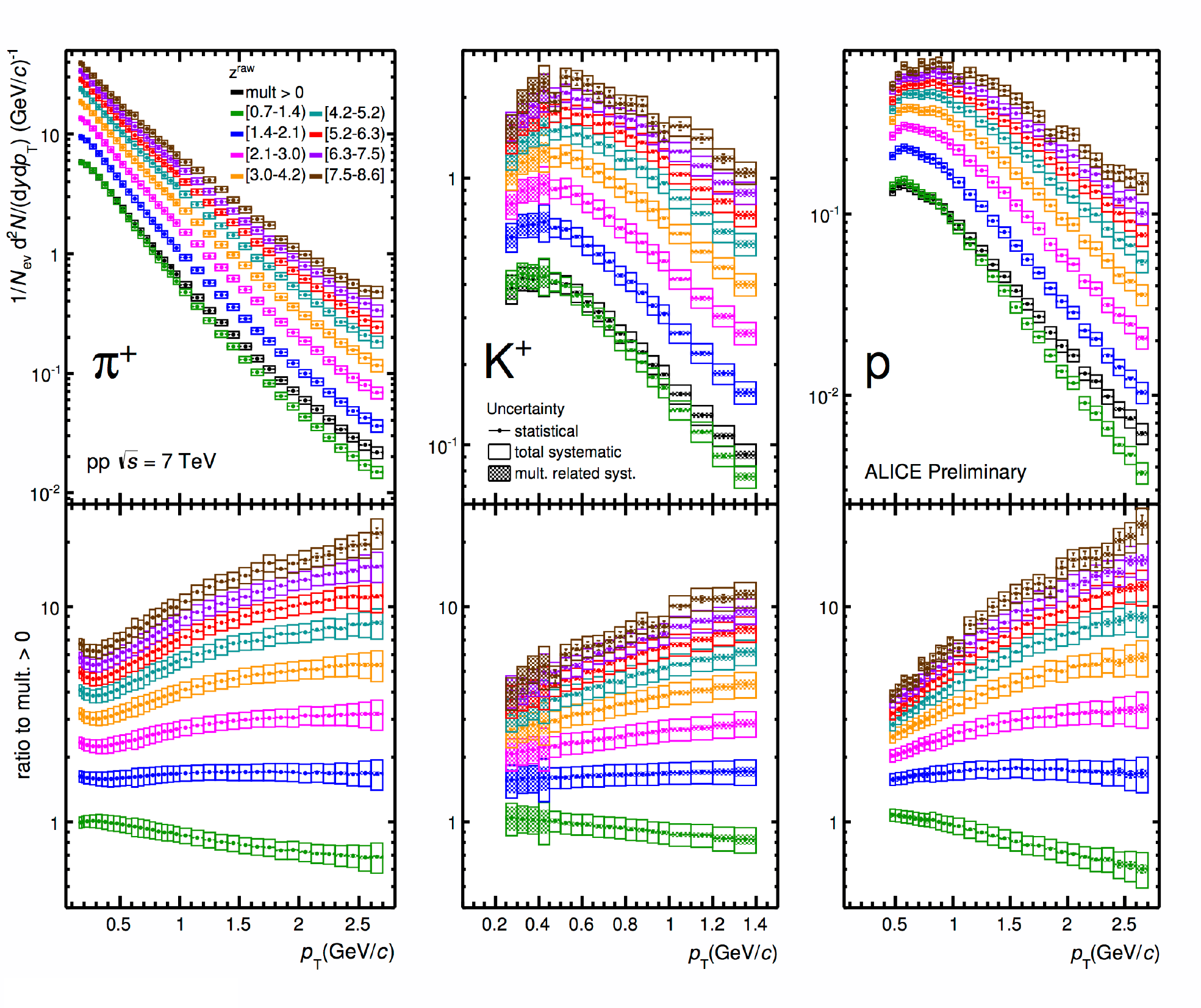}}
\subfloat[]{\includegraphics*[width=0.51\linewidth]{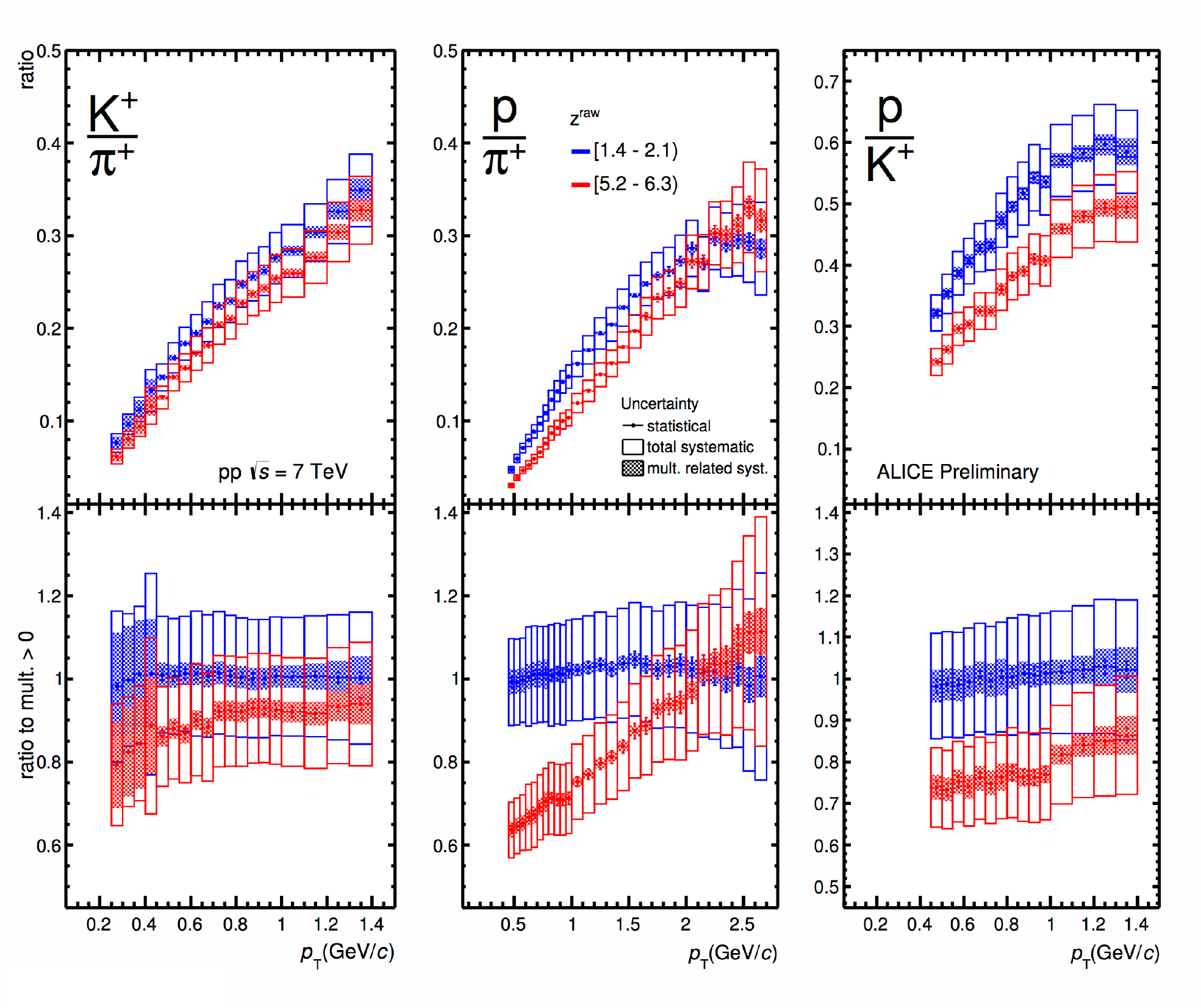}}
\caption{a) Upper row - charged particle multiplicity dependence of the transverse momentum distributions for \pip{}, \kap{} and p in pp collisions at 7 TeV; $z^{raw}$=$\left<N_{ch}^{raw}\right>_{mult\ bin}/\left<N_{ch}^{raw}\right>_{mult>0}$. 
Bottom row - ratio of transverse momentum distributions in a given multiplicity bin (z) relative to mult $>$ 0;
b) Upper row - \pt{} dependence of the particle ratios \kap{}/p, p/\pip{} and p/\kap{} in pp collisions at 7 TeV for two multiplicity bins. 
Bottom row - the ratio of the upper distributions relative to the one for mult $>$ 0.}
\label{fig:spectra}
\end{center}
\end{figure}
\noindent
All the correction factors were estimated using simulations based on the PYTHIA event generator \cite{16} (tune Perugia0 \cite{17}) and the GEANT3 \cite{18} transport code.
The corrections determined for the MB case were applied for all the multiplicity bins and their variations as a function of multiplicity, proved to be very small, were included in the systematic errors. 

The fully corrected \pt{} spectra for \pip{}, \kap{} and p were obtained by selecting events in eight multiplicity bins in the raw charged particle multiplicity distribution, 7-12, 13-19, 20-28, 29-39, 40-49, 50-59, 60-71 and 72-82 or, using the scaled multiplicity $z^{raw}$=$\left<N_{ch}^{raw}\right>_{mult \  bin}/\left<N_{ch}^{raw}\right>_{mult>0}$, [0.7-1.4), [1.4-2.1), [2.1-3.0), [3.0-4.2), [4.2-5.2), [5.2-6.3), [6.3-7.5) and [7.5-8.6]. 
The results are presented in Fig.~\ref{fig:spectra}a - upper row. 
In the bottom row of Fig.~\ref{fig:spectra}a the ratios of \pt{} distributions at different multiplicities relative to the mult \textgreater{} 0 case are represented. 
One could observe a systematic change in the shape of the ratios, i.e. a depletion in the low \pt{} region which shows a tendency to level off at larger \pt{} values. 
The amount of depletion clearly depends on the mass of the species and on the multiplicity, i.e. it is enhanced going from pions to protons and with increasing multiplicity for a given mass.
The \kap{}/\pip{},  p/\pip{} and p/\kap{} ratios as a function of \pt{} are plotted in Fig.~\ref{fig:spectra}b - upper row for the second and the sixth multiplicity bins. 
The general trends observed in Fig.~\ref{fig:spectra}a can be seen in a more quantitative way in these representations, especially in the bottom row of Fig.~\ref{fig:spectra}b, where the ratios of relative yields for the two multiplicities to the ones corresponding to $N_{ch}^{raw}$ \textgreater 0 are represented.
The observed depletion of p/\pip{} and p/\kap{} at low \pt{}, increasing with the multiplicity and decreasing towards 1.4-2.0 GeV/\textit{c} looks similar with the trends observed in A--A collisions, the heavier particles being pushed towards larger transverse momenta. 
Such behavior, observed in Pb--Pb collisions at 2.76 TeV \cite{ALICE-Pb}, was attributed to the existence of collective transverse flow.
The evolution of the \pt{} spectra shape with charged particle multiplicity in pp collisions at 7 TeV and  p--Pb at 5.02 TeV \cite{alice-pPb} and with centrality in Pb--Pb at 2.76 TeV \cite{ALICE-Pb} is rather similar.

\begin{figure}
\begin{center}
\subfloat[]{\includegraphics[height = 0.29\linewidth]{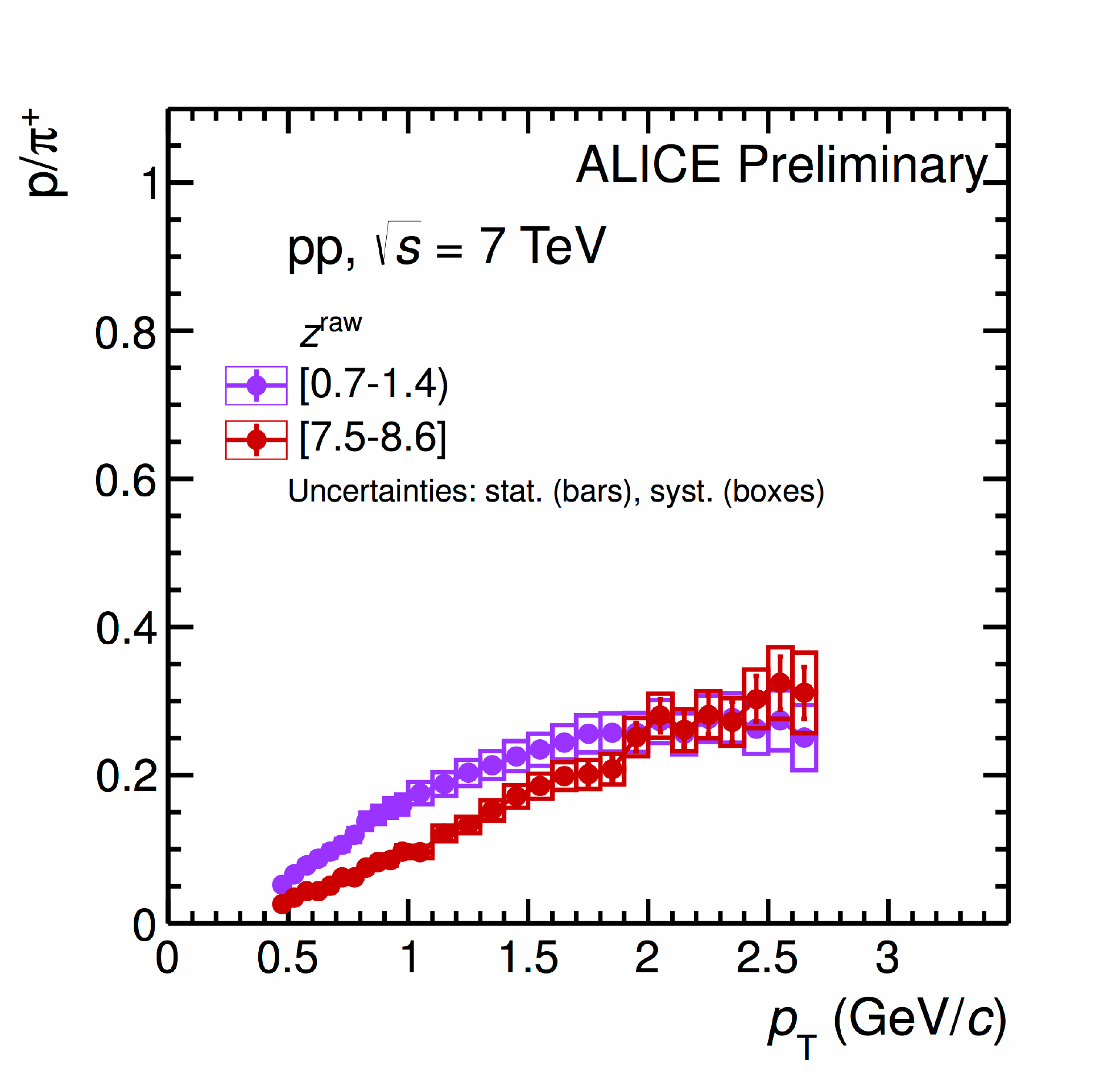}}
\subfloat[]{\includegraphics[height = 0.29\linewidth]{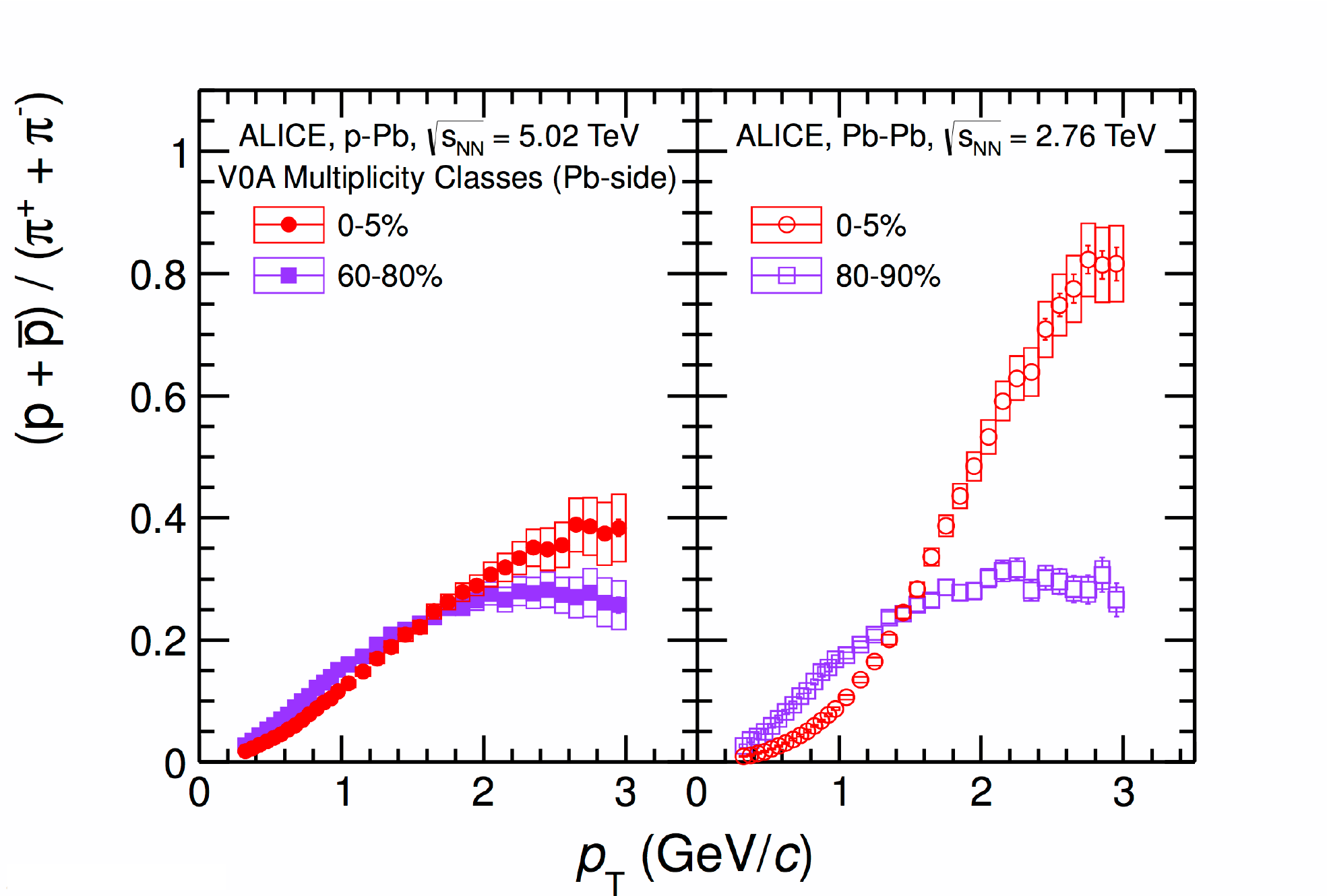}}
\subfloat[]{\includegraphics[height = 0.29\linewidth]{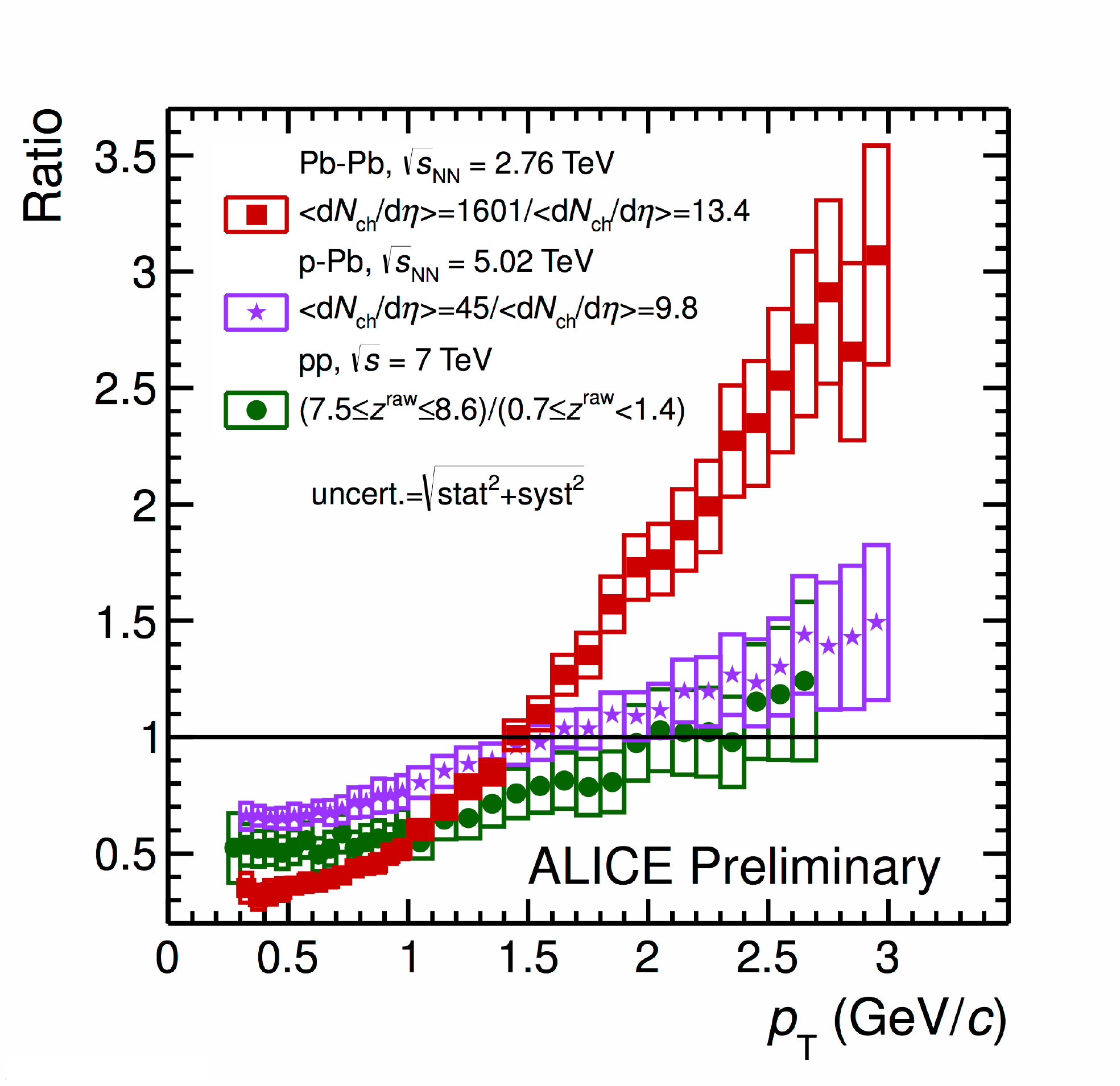}}
\caption{a) \pt{} dependence of p/\pip{} ratio for the second and highest multiplicity bins in pp collisions; 
b) \pt{} dependence of (p + \pbar{})/($\pi^{+}+\pi^-$) ratio in p--Pb for 60-80\% and 0-5\% multiplicity classes and Pb--Pb for 80-90\% and 0-5\% centrality \cite{alice-pPb};
c)~The ratios of the ratios presented in a) and b).}
\label{fig:systems_ratios}
\end{center}
\end{figure}
The \pt{} dependence of p/\pip{} for the second and highest multiplicity bins and of (p + \pbar{})/($\pi^++\pi^-$) ratio for p--Pb in 60-80\% and 0-5\% multiplicity classes and for Pb--Pb at 80-90\% and 0-5\% centrality \cite{alice-pPb}  are presented in Fig.~\ref{fig:systems_ratios}a and Fig.~\ref{fig:systems_ratios}b, respectively.
The push of protons towards larger \pt{} values relative to pions with increasing 
centrality or multiplicity is present for all three systems. 
Quantitatively, this can be followed in Fig.~\ref{fig:systems_ratios}c where the ratios of the ratios shown in Fig.~\ref{fig:systems_ratios}a and Fig.~\ref{fig:systems_ratios}b are presented. 
The ratio for pp follows closely the p--Pb trend as a function of \pt{}.

Based on these similarities one could investigate to what extent the quality and parameters of simultaneous fits of transverse momentum distributions for different multiplicities or centralities using expressions inspired by hydrodynamic models are also similar.
The quality of fits for \pip{}, \kap{} and p, based on the Boltzman-Gibbs Blast Wave (BGBW) expression \cite{Heinz}:
\begin{equation}
E\frac{d^3N}{dp^3}\sim
f(p_T) = \int_{0}^Rm_TK_1(m_Tcosh\rho/T_{kin})I_0(p_Tsinh\rho/T_{kin})rdr
\end{equation}
where $m_T=\sqrt{m^2+p_T^2}$; $\beta_r(r)=\beta_s(\frac{r}{R})^n$; $\rho=tanh^{-1}\beta_r$, is similar within the error bars for all three systems.
The \pt{} range on which the \pt{} spectra follows the hydro shape increases going towards higher multiplicity in pp and p--Pb  or higher centrality in Pb--Pb.

\begin{figure}
\begin{center}
\subfloat[]{\includegraphics*[height=0.375\linewidth]{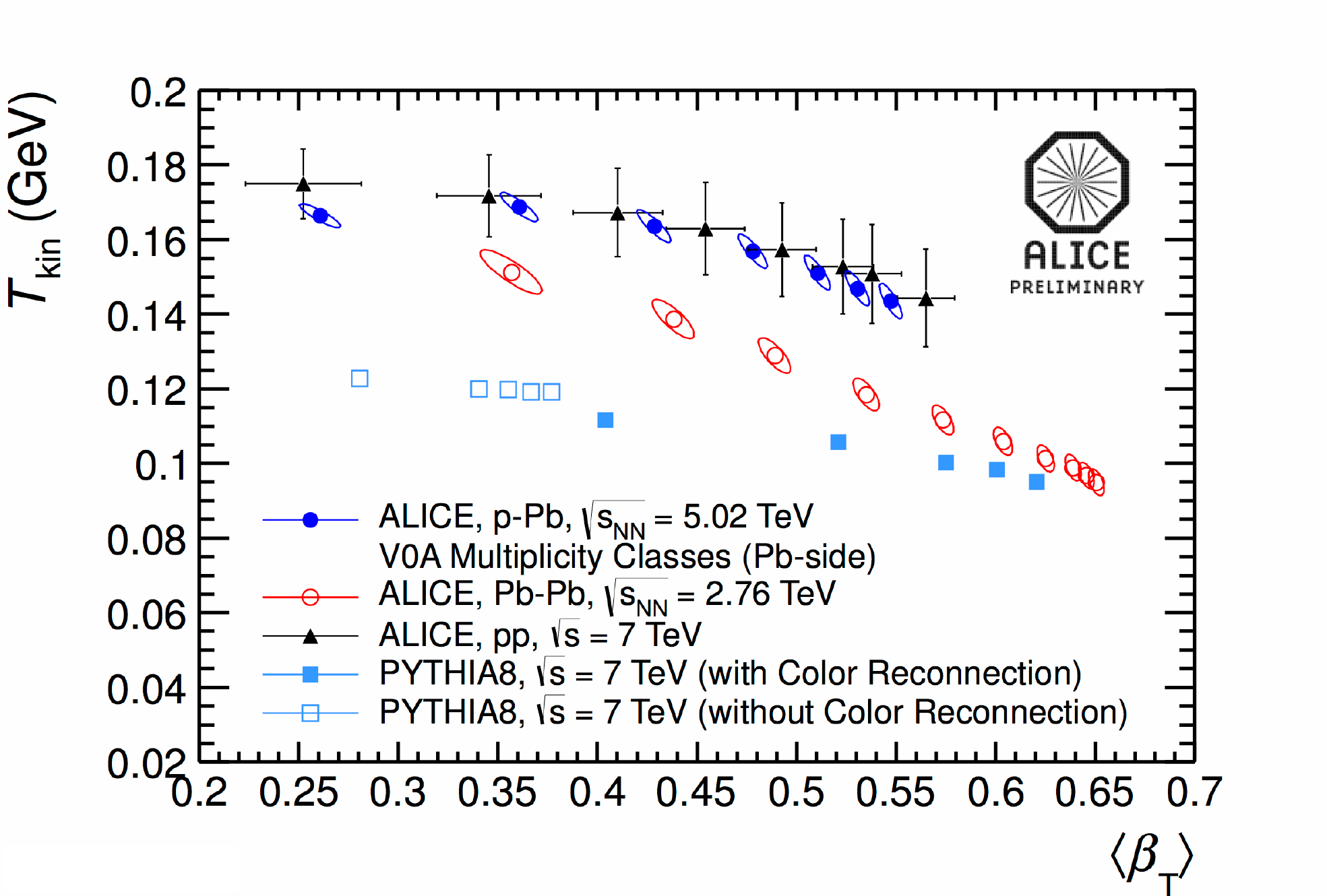}}
\subfloat[]{\includegraphics*[height=0.375\linewidth]{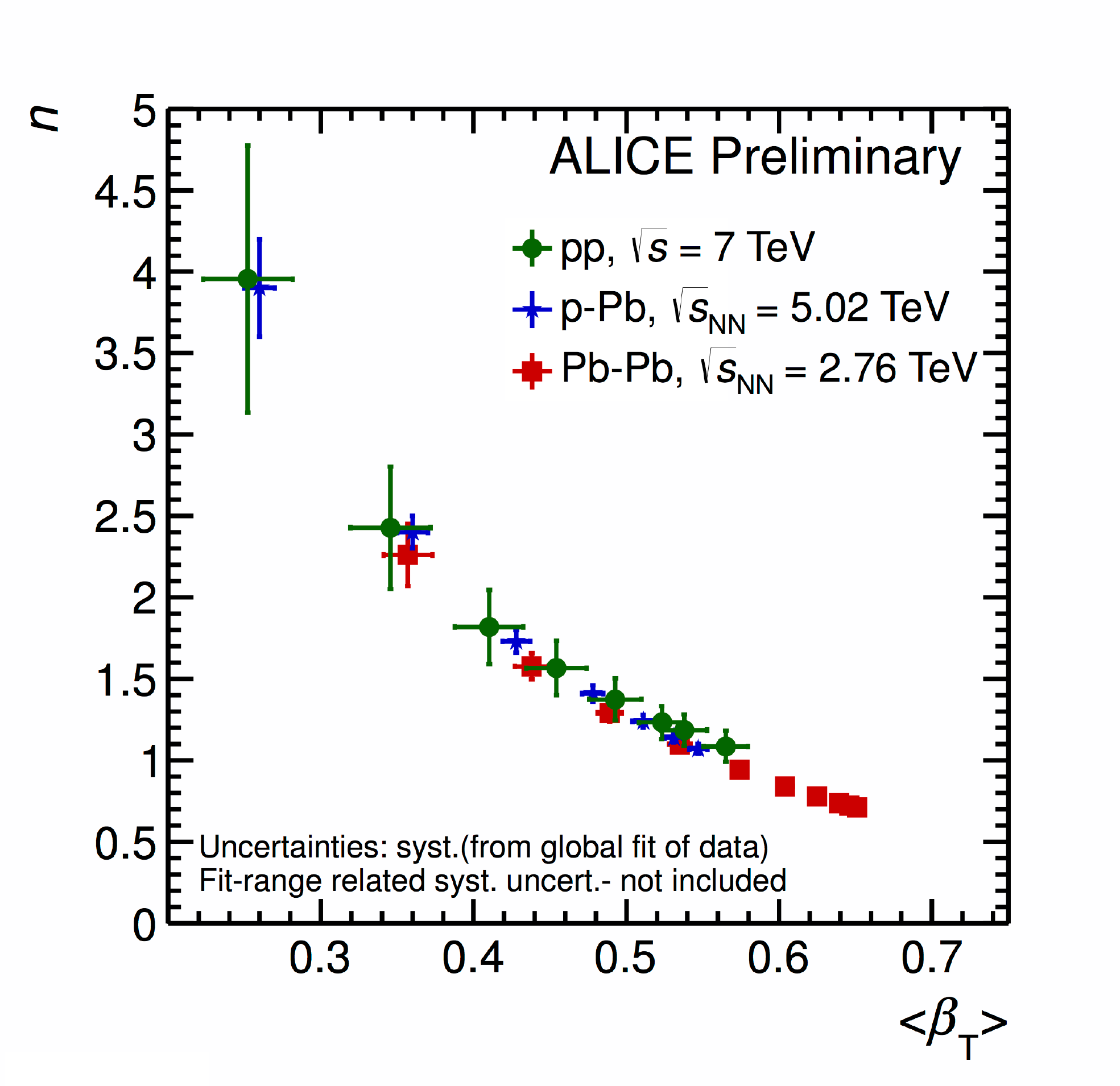}}
\caption{a) Freeze-out $T_{kin}$ versus transverse expansion velocity $\langle \beta_{\rm T} \rangle$ for different multiplicity bins for pp collisions at 7 TeV (black triangles) comparison with the results obtained for p--Pb (dark blue dots) for different multiplicity classes and Pb--Pb (red circles) as a function of centrality \cite{alice-pPb};
b) Expansion profile n versus transverse expansion velocity $\langle \beta_{\rm T} \rangle$ for different multiplicity bins for pp collisions at 7 TeV (green dots) compared with the results obtained for p--Pb (dark blue triangles) dots for different multiplicity classes and Pb--Pb (red squares) as a function of centrality \cite{alice-pPb}. The $ \langle \beta_{\rm T} \rangle$ values increase with multiplicity (pp and p--Pb) or centrality (Pb--Pb).}
\label{fig:BGBW_params}
\end{center}
\end{figure}

In Fig.~\ref{fig:BGBW_params}a the results of the fits for pp collisions at 7 TeV in terms of kinetic freeze-out temperature ($T_{kin}$) versus transverse expansion velocity ($\langle \beta_{T} \rangle$) are presented for different multiplicity bins and compared with the results obtained for p--Pb and Pb--Pb as a function of multiplicity classes and centrality, respectively \cite{alice-pPb}. 
While the $T_{kin}$-$\langle \beta_T \rangle$ correlation as a function of multiplicity in pp overlaps with the one corresponding to p--Pb as a function of multiplicity classes, for Pb--Pb the decrease of $T_{kin}$ and increase of $\langle \beta_T \rangle$ with centrality is more enhanced.
The observed trends are not reproduced by PYTHIA in absolute values although a somewhat similar trend seems to be present if Color Reconnection is included.

Further information accessed from these fits is the expansion profile, given by the parameter n.
The n-$\langle \beta_T \rangle$ correlation as a function of multiplicity or centrality for the three systems is presented in Fig.~\ref{fig:BGBW_params}b.
\noindent
Towards high charged particle multiplicity, the transverse expansion velocity approaches a linear dependence as a function of position in the fireball for pp, the trend overlapping with the one corresponding to p--Pb and Pb--Pb.

In conclusion, transverse momentum spectra of positive identified charged hadrons as a function of charged particle multiplicity in pp collisions at $\sqrt{s}$ = 7 TeV, up to multiplicity values never measured before, are reported. 
A clear depletion of p/\pip{} and p/\kap{} at low \pt{} , increasing with the multiplicity and decreasing towards larger \pt{}, similar with the trends observed in p--Pb and Pb--Pb collisions, is seen.
The kinetic freeze-out temperature ($T_{kin}$), expansion velocity ($\langle \beta_T \rangle$) and its profile extracted from the simultaneous fits of the \pip{}, \kap{} and p spectra with BGBW, show a multiplicity dependence trend similar with the ones obtained for p--Pb and Pb--Pb collisions as a function of multiplicity classes or centrality, respectively.
However, a conclusion about similar mechanisms for the three systems has to be taken with caution.
Detailed investigations based on theoretical approaches such as hydrodynamic models, parton based Gribov-Regge theory, Color Glass Condensate, Color Reconnection, will give insight to the underlying physics of this similar behavior observed at LHC energies.








\end{document}